\documentclass[aip,jcp,reprint,twocolumn,amsmath,amssymb]{revtex4-1} % J.Chem.Phys
\pdfoutput=1
\usepackage{amsmath,amsthm}
\usepackage{graphicx}% Include figure files
\usepackage{dcolumn}% Align table columns on decimal point
\usepackage{bm}% bold math
\usepackage{color}
\usepackage{epsfig}
\usepackage{multirow}
\usepackage{siunitx}% To typeset numbers and units
\usepackage{subfig}% for subfigures
\usepackage{todonotes}% For TODO-boxes, all inlined
\makeatletter
\presetkeys
    {todonotes}
    {inline}{}
\makeatother

\graphicspath{{figs/}}

\newcommand{\Vect}[1]{\ensuremath{\bm{#1}}} % bold vector symbols
\newcommand{\model}[1]{{\bf #1}}

\begin{document}

\preprint{Preprint}

\title{Simulation of Electric Double Layers around Charged Colloids \\ in Aqueous Solution of Variable Permittivity}

\author{Florian \surname{Fahrenberger}$^1$}%\email{floh@icp.uni-stuttgart.de}
\author{Zhenli \surname{Xu}$^2$}%\email{zhenli@icp.uni-stuttgart.de}
\author{Christian \surname{Holm}$^1$} \email{holm@icp.uni-stuttgart.de}

\affiliation{1. Institute for Computational Physics, University of Stuttgart, Stuttgart 70569, Germany}
\affiliation{2. Department of Mathematics, Institute of Natural Sciences and MoE Key Lab of Scientific and Engineering Computing, Shanghai Jiao Tong University, Shanghai 200240, China}

%%Lines break automatically or can be forced with \\

\date{\today}

\begin{abstract}
The ion distribution around charged colloids in solution has been investigated intensely during the last decade. However, few theoretical approaches have included the influence of variation in the dielectric permittivity within the system, let alone in the surrounding solvent. In this article, we introduce two relatively new methods that can solve the Poisson equation for systems with varying permittivity. The harmonic interpolation method (HIM) approximately solves the Green's function in terms of a spherical harmonics series, and thus provides analytical ion-ion potentials for the Hamiltonian of charged systems. The Maxwell Equations Molecular Dynamics (MEMD) algorithm features a local approach to electrostatics, allowing for arbitrary local changes of the dielectric constant. We show that the results of both methods are in very good agreement. We also found that the renormalized charge of the colloid, and with it the effective far field interaction, significantly changes if the dielectric properties within the vicinity of the colloid are changed.
\end{abstract}

%%%%% AMS/PACs/Keywords %%%%%%%%%%%
%\pacs{ }
%\ams{}
%\keywords{ }

%%%% maketitle %%%%%
\maketitle

\section{Introduction}

When a charged nanoparticle (colloid or biopolymer) is immersed in an aqueous solution, counterions accumulate near the interface between the macroion and the solvent, forming an ionic cloud which neutralizes the bare charge. The structure of this ionic cloud is called the electric double layer (EDL) and plays an important role in the physical and chemical properties of many systems at small scales.~\cite{levin02a,grosberg02a,french10a,walker11a,zwanikken13a,degraaf08a,rekvig04a,rekvig07a} The ion distributions in EDLs are the result of the balance of electrostatic interactions and entropic repulsion, and are affected by various factors such as surface charge density and discreteness, ionic size and valency, and the system temperature.\cite{french10a,walker11a}

In response to the already present, widely spread, experimental data, researchers are increasingly examining the properties of colloids from a simulational approach.\cite{evans99a,arora98a,kreer06a}  The use of coarse grained models is crucial for dealing with systems on such a big scale. In addition to reducing the colloid to one large sphere or sometimes treating it as an infinite plane, \cite{semenov13a}  one of the most common coarse graining approaches is to treat the solvent particles in the colloidal suspension on a continuum level. This includes the use of implicit fluid solvers, e.g., the Lattice-Boltzmann method, and in addition the introduction of a bulk dielectric permittivity to account for the polarizability of the water molecules.

With the introduction of techniques such as the induced charge computation method,\cite{boda04b,tyagi10a,kesselheim11a}  extended Poisson-Boltzmann solvers,\cite{ipbs-github} and other functional approaches,\cite{jadhao12a,jadhao13a,bichoutskaia10a} it has become possible to model a dielectric jump at the surface of the colloid, setting the dielectric permittivity of the colloid to a more realistic number $\varepsilon_C=2$, and that of the surrounding water solvent to $\varepsilon_W=80$. It has been shown that this dielectric jump even has a major influence on the far field electrostatic potential of the colloid, and therefore should not be neglected in simulations.\cite{messina01a, jadhao12a, jadhao13a, dossantos11a}

While this sharp dielectric contrast is already closer to the physical behavior of the system, it is still far from an accurate description. It has been shown that, in close proximity to a highly charged surface, solvent molecules align and form structures, reducing the strength of dielectric displacement in this region. This effect is enhanced by the presence more salt ions in the EDL, further decreasing the flexibility of the solvent dipoles and therefore the polarizability.\cite{bonthuis11a,bonthuis12a} While the influence of the dielectric mismatch between solute and solvent has been studied,\cite{torrie82a,torrie84a,kjellander84a,wang09j,dossantos11a,gan12a} research on dielectric changes within the EDL itself is still relatively new. In general, Poisson's equation with a space-dependent coefficient can only be solved by grid-based finite difference or finite element methods.~\cite{im98a,lu08b} These methods are computationally expensive if used in particle-based computer simulations, since the equation has to be solved billions of times per simulation.

In this article, we present two methods capable of dealing with spatially varying dielectric permittivities. One is the harmonic interpolation method (HIM) by extending the three-layer model of biomolecular solvation,~\cite{qin09a,xue13a} which uses an analytical-based Poisson solver to include an environment of spherical-symmetric dielectric functions within the EDL. The other is an extension~\cite{fahrenberger13a-pre} of the Maxwell Equations Molecular Dynamics (MEMD)~\cite{pasichnyk04a} algorithm to include arbitrary changes of the dielectric constant interpolated on a lattice. Using these two algorithms, we study the influence of different dielectric functions (schematically illustrated in Fig.~\ref{fig:model-functions}) on the EDL structure of colloidal systems in solution. We compare the results from the two entirely different algorithms to ensure the correctness of our findings. The results are then analyzed via the radial distribution function (RDF) and charge renormalization~\cite{alexander84a}  which is useful for providing effective charges in the DLVO theory,\cite{derjaguin41a,verwey48a} and it is shown that the EDL structure and the effective charge of a colloid is modified significantly by the dielectric environment near the interface. These simulation results also demonstrate the usefulness of the developed algorithms.

In what follows, the theory of the EDL model will be first explained, and the two electrostatics algorithms, HIM and MEMD, will be introduced. Next, the simulation setup will be introduced, and finally, the results of our simulations using both approaches will be presented, compared, and discussed.

 %%%% Start %%%%%%

\section{Electric double layer model}

\begin{figure}[htbp!]
\subfloat[Model][Schematic of the system]{
    \centering
    \includegraphics[width=.8\linewidth]{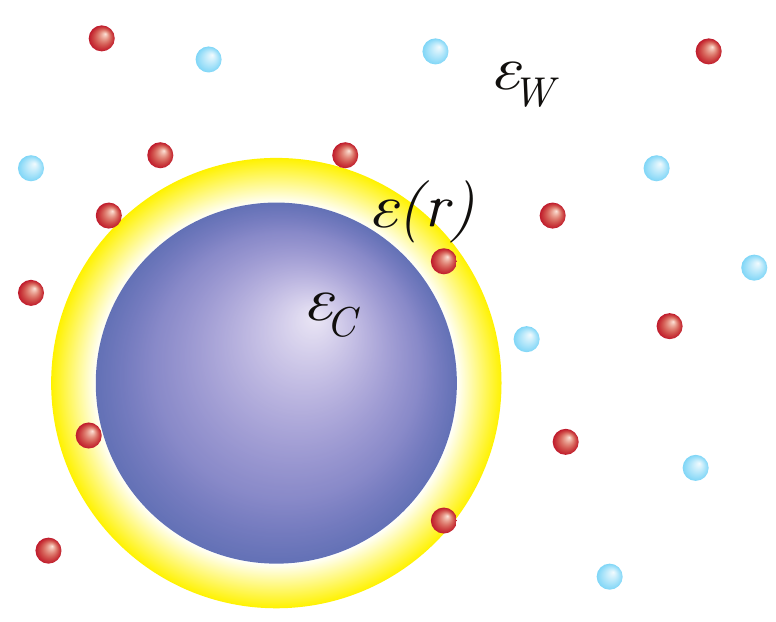}
    \label{fig:colloid-schematic}
}

\subfloat[Model][Permittivity functions]{
    \centering
    \includegraphics[width=.9\linewidth]{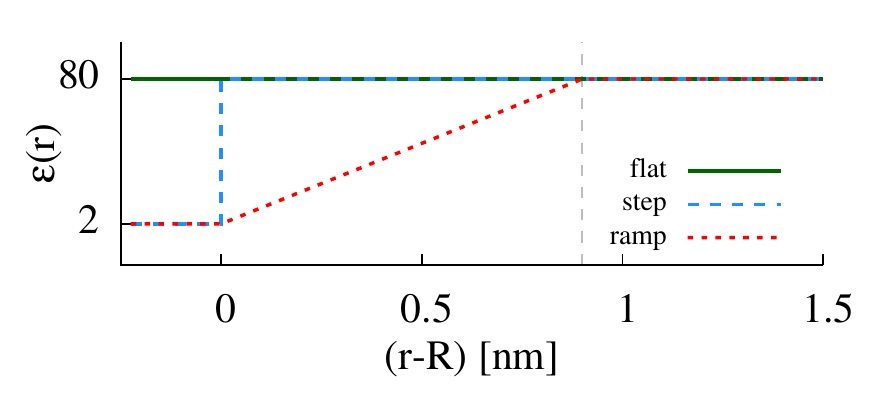}
    \label{fig:model-functions}
}
\caption{(a) The simulation setup: A charged colloid is suspended in a solvent with surrounding counterions and salt ions. The system's dielectric permittivity is divided into three regions: $\varepsilon_C$ within the colloid, $\varepsilon_W$ in bulk water, and $\varepsilon(r)$ in the intermediate region. (b) Three different models are investigated: The first model \model{flat} is the most simplistic, and often used, approach with a constant background permittivity $\varepsilon_C = \varepsilon_W = \varepsilon(r) = 80$ throughout the system. The second model \model{step} introduces a sharp contrast in form of a step function with $\varepsilon_C=2$ within the colloid, and $\varepsilon(r) = \varepsilon_W = 80$ outside. The third model \model{ramp} linearly interpolates between the two regimes $\varepsilon_C=2$ in the colloid and $\varepsilon_W=80$ in bulk water. The interpolation occurs over $2$ ion diameters, which means that the ions can enter the region of lowered dielectric permittivity.} \label{fig:system-setup}
\end{figure}

Theoretical descriptions of EDLs of charged colloids in electrolytes are often based on the primitive model.~\cite{mcmillan45a,linse05a} In this model, ions of different species are represented by charged hard spheres, differing in size and valency (Fig.~\ref{fig:colloid-schematic}). The ions interact with each other by a combination of a short-range hard-sphere potential and a long-range electrostatic potential in a solvent which is described by its macroscopic dielectric permittivity. The equilibrium-state properties of the EDLs can be calculated by integral equation theories or Molecular Dynamics/Monte Carlo (MD/MC) computer simulations with the Hamiltonian composed of ion-ion and ion-interface interactions, and appropriate boundary conditions.

We consider a colloidal sphere of radius $R$ with surrounding electrolyte, shown in Fig.~\ref{fig:colloid-schematic}. The dielectric permittivity is a function of radial distance, $\varepsilon(r)$, which takes a constant $\varepsilon_C$ within the sphere $r\leq R$, another constant $\varepsilon_W$ in the bulk solvent, and depends on $r$ in the intermediate region. We study the profiles shown in Fig.~\ref{fig:model-functions}. For a source point charge $q_s=1$ at $\mathbf{r}_s$, the potential due to this charge, called the Green's function $G(\Vect{r},\Vect{r}_s)$, is the solution of the following variable coefficient Poisson's equation,
\begin{equation}
-\nabla\cdot\varepsilon(r)\nabla G(\mathbf{r},\mathbf{r}_s)=4\pi  \delta(\mathbf{r}-\mathbf{r}_s),
\label{eq:poisson}
\end{equation}
where $\delta$ is the Dirac delta, and the divergence $\nabla\cdot$ and the gradient $\nabla$ are both with respect to coordinates $\mathbf{r}$. The Green's function is a function of the source point $\mathbf{r}_s$ and the field point $\mathbf{r}$, thus living on a six dimensional domain, and can therefore not be efficiently solved by numerical methods.

As will be discussed in the next section, the solution of the Green's function in a sphere-symmetric dielectric medium  can be written into a sum of a multipole series and a direct Coulomb interaction,
\begin{equation}
G(\mathbf{r}_i,\mathbf{r}_j)=G_\mathrm{pol}(\mathbf{r}_i,\mathbf{r}_j)+G_\mathrm{coul}(\mathbf{r}_i,\mathbf{r}_j), \label{green}
\end{equation}
where $G_\mathrm{pol}$ and $G_\mathrm{coul}$ correspond to the series term and the Coulomb term shown in Eq.~\eqref{potential} below, respectively.
When the Green's function is given, the electrostatic interaction between any two charges $q_i=Z_ie$ and $q_j=Z_je$, with valencies $Z_i$ and $Z_j$, can be expressed as
\begin{equation}
U_{ij}=\frac{e^2}{4\pi\varepsilon_0}Z_iZ_j G(\mathbf{r}_i,\mathbf{r}_j),
\end{equation}
where $e$ is the unit charge of an electron and $\varepsilon_0$ is the vacuum dielectric permittivity.
Besides pairwise interactions, mobile ions have self energy fluctuations in an inhomogeneous dielectric medium. The second term $G_\mathrm{coul}$ in Eq.~\eqref{green} is divergent for the self Green's function with overlapping source and field points, which is invariant and can be absorbed into the chemical potential for uniform media. This term cannot be discarded when the medium has a varying dielectric permittivity as the solvation energy changes with it.
The direct Coulomb term could be substituted by a modification of the Born energy~\cite{born20a} to introduce the local self energy contribution, and the self energy of ion $i$ is given by
\begin{equation}
U_i^\mathrm{self}= \frac{Z_i^2 e^2}{4\pi\varepsilon_0} \left[ \frac{G_\mathrm{pol}(\mathbf{r}_i,\mathbf{r}_i)}{2}
+\frac{1}{2\varepsilon(r_i)a_i} \right],
\label{eq:solvation-energy}
\end{equation}
where $a_i$ is the Born radius of the ion, which assumes the value of the ion radius in our simulations. The modified Born energy is a local contribution due to the finite size of the ion, and the  $G_\mathrm{pol}$ term is a global contribution due to the dielectric variation. The splitting of the self energy can be derived by a perturbation expansion, since that the Born radius is treated as a small parameter~\cite{wang10d}. The Hamiltonian of the system is then given by, $U=U_\mathrm{hs}+U_\mathrm{elec}$, where
\begin{equation}
U_\mathrm{elec}= \sum_{i<j}U_{ij}+\sum_{i=1}^N U_i^\mathrm{self}.\label{hamiltonian}
\end{equation}
The hard-sphere potential $U_\mathrm{hs}$ is the infinite repulsion when any two spherical particles have an overlap, and otherwise does not contribute to the Hamiltonian.

One characteristic length for electrolytes is the Bjerrum length $\ell_B=\beta e^2/(4\pi\varepsilon_0\varepsilon_W)$, where $\varepsilon_0$ is the vacuum permittivity,  $\varepsilon_W$ is the relative permittivity of the bulk solvent, and $\beta=1/k_BT$ is the inverse thermal energy. The Bjerrum length is the distance between two unit charges at which they interact with the thermal energy~$k_BT$. At room temperature, the Bjerrum length of bulk water solvent is $\ell_B=\SI{0.714}{nm}$.

\section{Harmonic interpolation method for Green's function}

The Green's function~\eqref{green} can be used to investigate the physics of charged systems through particle-based computer simulations (MC or MD), which are often limited due to the lack of an efficient Poisson solver. Only limited analytical Green's function solutions exist such as for a planar or spherical interface separating two media of constant permittivity.~\cite{jackson99}
When $\varepsilon(r)$ is constant,  it is known that the Green's function is the reciprocal distance divided by the dielectric, $1/(\varepsilon|\mathbf{r}-\mathbf{r}_s|)$. If the solute  and solvent medium have a sharp dielectric mismatch,  characterized by $\varepsilon_C$ and $\varepsilon_W$, respectively, the solution is obtained by spherical harmonics~\cite{jackson99} or image charges.~\cite{neumann83c,lindell93a} For a general $\varepsilon(r)$ the Green's function is not exactly solvable. However, in the following, we will develop the harmonic interpolation method to obtain accurate approximate solutions using an almost-analytical expression, by extending the algorithm for the three-layer model of biomolecular solvation. \cite{qin09a}

To solve Eq.~\eqref{green}, we divide the radial distance into $L$ intervals with $L+1$ points $r_0, r_1,\cdots, r_L$ and denote the $l$th interval $(r_{l-1},r_l)$ by $I_l$. This division could be general, but in our problem we choose the first interval to cover the region in the sphere, i.e., $r_0=0$ and $r_1=R$. We approximate the dielectric function by piecewise functions,
$\varepsilon(r)\approx\varepsilon_l(r)$, for $r\in I_l$. In physically realistic systems, the transition layer from $\varepsilon_C$ to $\varepsilon_W$ is less than 1 nanometer in width. It therefore usually suffices to work with a small value of $L$. The permittivity in the $l$th interval is approximated by,
\begin{equation}
\varepsilon_l(r)=\left(a_l+\frac{b_l}{r}\right)^2,
\end{equation}
 where the coefficients $a_l$ and $b_l$ are interpolated from the values $\varepsilon(r_{l-1})$ and $\varepsilon(r_l)$ at two interval ends. Clearly we have $\nabla^2\sqrt{\varepsilon_l(r)}=0$, because the square root of the permittivity in each interval is harmonic.

Suppose the source charge resides in the $k$th interval, $r_s\in I_k$. The solution in each layer is rewritten as,
\begin{equation}
G(\mathbf{r},\mathbf{r}_s)=\Phi_l(\mathbf{r}),~~\hbox{for}~~r\in I_l.
\end{equation}
Here the electric potential at $l$th layer, $\Phi_l(\mathbf{r})$, should be also the function of the source position, but we dropped the dependence for notational simplicity.
Since $\sqrt{\varepsilon_l(r)}$ is harmonic, by a simple derivation, the potential satisfies,~\cite{qin09a}
\begin{equation}
-\sqrt{\varepsilon_l(r)}\nabla^2\left[\sqrt{\varepsilon_l(r)}\Phi_l(\mathbf{r})\right] = 4\pi  \delta(\mathbf{r}-\mathbf{r}_s).
\end{equation}
Since $r_s\in I_k$, we can write $\delta(\mathbf{r}-\mathbf{r}_s)/\sqrt{\varepsilon_l(r)}=\delta(\mathbf{r}-\mathbf{r}_s)/\sqrt{\varepsilon_k(r_s)}.$
The equation becomes a constant coefficient Poisson's equation for $\sqrt{\varepsilon_l(r)}\Phi_l(\mathbf{r})$,
\begin{equation}
-\nabla^2\left[\sqrt{\varepsilon_l(r)}\Phi_l(\mathbf{r})\right] = \frac{4\pi  \delta(\mathbf{r}-\mathbf{r}_s)}{\sqrt{\varepsilon_k(r_s)}}.
\end{equation}
The potential function can then be expanded in terms of spherical harmonics series,
\begin{eqnarray}
&\Phi_l(\mathbf{r})&=\frac{1}{\sqrt{\varepsilon_l(r)}}\sum_{n=0}^\infty\left[
A_l(n)r^n+B_l(n)r^{-n-1}\right]P_n(\cos\theta) \nonumber\\
&&~~~+\frac{\delta_{lk} }{\sqrt{\varepsilon_l(r)\varepsilon_k(r_s)}|\mathbf{r}-\mathbf{r}_s|}, \label{potential}
\end{eqnarray}
where $\delta_{lk}$ is the Kronecker delta, $\theta$ is the angle between $\mathbf{r}$ and $\mathbf{r}_s$, and $P_n(\cdot)$ is the Legendre polynomial of order $n$. We have $B_1(n)=0$ and $A_L(n)=0$ since the potential is finite in the innermost and the outermost layers, and other coefficients $A_l(n)$ and $B_l(n)$ are determined by the continuity conditions on each layer boundary,
\begin{equation}\left\{\begin{array}{ll}
\Phi_l(\mathbf{r}_l)=\Phi_{l+1}(\mathbf{r}_l),~~\hbox{and}  \\
\varepsilon_l(r_l)\partial_r\Phi_l(\mathbf{r}_l)=\varepsilon_{l+1}(r_l)\partial_r\Phi_{l+1}(\mathbf{r}_l),
\end{array}
\right.\end{equation}
for $l=1,\cdots, L-1.$ In the case of the dielectric function being continuous, we have $\varepsilon_l(r_l)=\varepsilon_{l+1}(r_l)$.

We need to introduce the spherical harmonics expansion of the reciprocal distance,
\begin{equation}
\frac{1}{|\mathbf{r}-\mathbf{r}_s|}=
\sum_{n=0}^\infty \frac{r_<^n}{r_>^{n+1}}P_n(\cos\theta),
\end{equation}
where $r_<$ and $r_>$ are ordered to obey $(r_<) < (r_>)$ for $r$ and $r_s$, respectively. Then the potential can be written in a compact form,
\begin{equation}
\Phi_l(\mathbf{r})=\sum_{n=0}^\infty M_{l,n}(r) P_n(\cos\theta),
\end{equation}
with
\begin{equation}
M_{l,n}(r)=\frac{A_l(n)r^{2n+1}+B_l(n)}{\sqrt{\varepsilon_l(r)}r^{n+1}}+
\frac{\delta_{lk}}{\sqrt{\varepsilon_l(r)\varepsilon_k(r_s)}}\cdot\frac{r_<^n}{r_>^{n+1}}. \label{Mln}
\end{equation}

The spherical harmonics are orthogonal, which means for each $n$ the boundary conditions lead to two equations for $A_l(n)$ and $B_l(n)$ at each interface between two layers. This gives,
\begin{equation}\left\{\begin{array}{ll}
M_{l,n}(r_l)=M_{l+1,n}(r_l),  \\
\varepsilon_l(r_l)\partial_rM_{l,n}(r_l)=\varepsilon_{l+1}(r_l)\partial_rM_{l+1,n}(r_l),
\end{array}
\right. \label{bc}\end{equation}
for $l=1,\cdots, L-1.$ There are $2(L-1)$ equations for $2(L-1)$ unknowns. We thus obtain the following system of linear equations, $\mathcal{M}_n\mathbf{b}_n=\mathbf{f}_n$, where $\mathcal{M}_n$ is the coefficient matrix for the $n$th multipole term, $\mathbf{b}_n=(A_1(n), A_2(n), B_2(n),\cdots,A_{L-1}(n), B_{L-1}(n), B_L(n))^T$ with $B_1(n)=A_L(n)=0$, and $\mathbf{f}_n$ are from the source contribution, the second term of Eq.~\eqref{Mln}.

Note that the matrix $\mathcal{M}_n$ is independent of the source point $\mathbf{r}_s$, and its inverse can be calculated before the running of simulations, i.e., in each step, the solution is $\mathbf{b}_n=\mathcal{M}_n^{-1}\mathbf{f}_n$ where $\mathcal{M}_n^{-1}$ is given explicitly by the Gaussian elimination and stored in memory throughout all calculations. Therefore, only operations of matrix-vector multiplication are required when the method is used in Monte Carlo simulations, and the number of operations is the order of $(L-1)^2$ for the computation of the coefficients for each $n$.

\section{Maxwell Equations Molecular Dynamics}

Another approach for spatially varying dielectric permittivity is the relatively new electrostatics method, the Maxwell Equations Molecular Dynamics (MEMD). The idea was first introduced by Maggs and Rossetto \cite{maggs02a} in 2002 and later extended to wavelike propagation of the magnetic field component for Molecular Dynamics (MD) simulations by Rottler and Maggs~\cite{rottler04a,rottler04b}, and parallel by D\"{u}nweg and Pasychnik.\cite{pasichnyk04a}

In this algorithm, instead of solving the global electrostatic potential via the Poisson equation~\eqref{eq:poisson}, the electric field is derived directly from the time derivative of Gauss' law $\nabla\Vect{D}=\rho$ of electrodynamics, with $\Vect{D}=\varepsilon(\Vect{r})\Vect{E}$. Additionally, an initial solution of the Poisson equation is calculated and updated each time step. This integrated equation offers an additional degree of freedom in the form of an arbitrary conservative vector field $\Vect{\Theta}$. In its most general form it reads
\begin{equation}
\dot{\Vect{D}} + \Vect{j} - \nabla\times\dot{\Vect{\Theta}} = 0,
\label{eq:memd-constraint}
\end{equation}
where $\Vect{j}$ denotes the local electric current. This constraint can be enforced via a Lagrange multiplier $\Vect{A}$. Using variational calculus, this naturally leads to the Maxwell equations of motion for the charges and fields. The degree of freedom introduced in~\eqref{eq:memd-constraint} is fixed using the Weyl gauge
\begin{equation}
\dot{\Vect{A}} = -\frac{\Vect{D}}{\varepsilon},\label{eq:weyl-gauge}
\end{equation}
and we introduce the magnetic field
\begin{equation}
\Vect{B} \mathrel{\mathop:}= \nabla\times\Vect{A} .
\end{equation}
The actual electrostatic potential $\Phi$ is never calculated in this algorithm, only the electric field $\Vect{E}$ for calculating the Lorentz force
\begin{equation}
\Vect{F}=q\cdot(\Vect{E}+\Vect{v}\times\Vect{B}) .
\label{eq:lorentz-force}
\end{equation}

Simply by applying the constraint~\eqref{eq:memd-constraint} and the Weyl gauge~\eqref{eq:weyl-gauge}, the complete set of equations of the electromagnetic formalism is reproduced. It has also been shown that the propagation speed of the magnetic field, an equivalent of the speed of light $c$, can be reduced in a Car-Parrinello manner, and correct retarded solutions for statistic observables are maintained.\cite{maggs02a}

This reduces the elliptic partial differential equation~\eqref{eq:poisson} to a set of hyperbolic differential equations for the propagation of magnetic fields and charges, requiring only local operations for the solution. It therefore opens the possibility of arbitrary local dielectric permittivities within the system. If discretized on a lattice, and coupled with a linear next neighbor interpolation scheme for the charges and electric currents, the permittivity can be set individually on every lattice link. This is in agreement with $\varepsilon(\Vect{r})$ being a differential 1-form, if we assume the tensor to only have identical diagonal entries (optically isotropic dielectric medium).

\begin{figure}[htbp!]
    \centering
    \includegraphics[width=.7\linewidth]{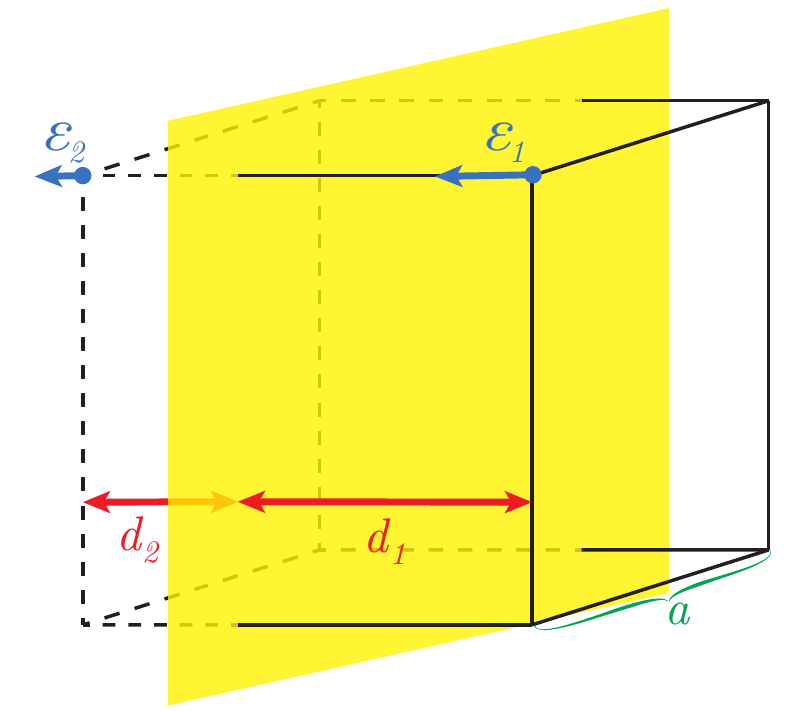}
    \caption{Interpolation of dielectric permittivity values on the lattice. $\varepsilon(\Vect{r})$ has a position and a direction (blue arrow). The values for $\varepsilon_1$ and $\varepsilon_2$ are determined and the value on the connecting link is set to the average value. If the gradient is too large, the link is marked as \emph{interface link}. The values for interface links are determined by linearly interpolating the sharp dielectric interface according to its position.}
    \label{fig:memd-epsilon-interpolation}
\end{figure}

The values for $\varepsilon(\Vect{r})$ on the links are determined as depicted in Fig.~\ref{fig:memd-epsilon-interpolation}. The values for two adjacent lattice sites are averaged if their difference is less than 10 percent of their value, i.e., $2|\varepsilon_1 - \varepsilon_2|/(\varepsilon_1 + \varepsilon_2)\le 0.1$. If the difference is larger, the link is marked as an \emph{interface link}, meaning that is passes through a sharp interface. In a second pass, the exact intersection of these sharp interfaces with the marked links is determined, and the value is interpolated linearly
\begin{equation}
\varepsilon_\text{link} = \varepsilon_1\cdot\frac{d_2}{a} + \varepsilon_2\cdot\frac{d_1}{a},
\label{eq:dielectric-interpolation}
\end{equation}
where $\varepsilon_1$ and $\varepsilon_2$ are the permittivity values on the adjacent lattice sites on each side of the interface respectively, $d_1$ and $d_2$ are the distances of the according lattice site along the link to the interface, and $a$ is the lattice spacing, as depicted in Fig.~\ref{fig:memd-epsilon-interpolation}.

In this algorithm, the charges can move freely through a smoothly varying dielectric medium. The numerical error depends on the finite propagation speed of the magnetic fields and the coarseness of the linear interpolation scheme. A relative force error of $10^{-3}$ however is achievable in sufficiently homogenous systems.

\section{Simulation setup}

We have employed canonical ensemble (NVT) Monte Carlo simulations with the HIM for electrostatics and the Metropolis algorithm,\cite{metropolis53a,frenkel02b} and equilibrium Molecular Dynamics simulations for charged systems in different media. A spherical colloid of radius $R$ is placed at the center of a simulation volume which is filled with an electrolyte solution described by the primitive model of constant dielectric permittivity (see Fig.~\ref{fig:colloid-schematic}). The bare charge of the colloid $Q_C=Z_C\text{e}$ is assumed to be uniformly distributed on the surface, and is equivalently placed at the center by Gauss' law.  Initially, small ions are randomly distributed in the solvent. Each ion brings a charge $q_i=Z_ie$ where $Z_i=\pm Z$ and is modeled by a hard sphere of diameter $D_\text{ion}$. Overall the system is electrically neutral, $(N_+-N_-)Z+Z_C=0$. In all calculations, we take  $R=\SI{4}{nm}$,  $D_\text{ion}=\SI{0.45}{nm}$ and a large cell radius $R_\mathrm{cell}=\SI{10}{nm}$ to reduce boundary effects. The simulations are performed at room temperature.

The cell in the MC simulations is modeled by a Wigner-Seitz (WS) spherical cell of radius $R_\mathrm{cell}=\SI{10}{nm}$, with hard wall boundary conditions. The MD simulation cell is cubic with box length $\SI{20}{nm}$ and has periodic boundaries. We chose the cell size intentionally big compared to the Debye length of the system (between $\num{1}$ and $\SI{2}{nm}$ for our salt concentrations) to avoid any influence of the difference in periodicity between HIM and MEMD.

The following three models of different dielectric environments are adopted for comparison, corresponding to Fig.~\ref{fig:model-functions}. The function is separated into three spherical regions, $\varepsilon_C$ within the colloid, $\varepsilon_W$ in the bulk, and $\varepsilon(r)$ within the solvent close to the surface.

Model~1 (\model{flat}) uses a homogeneous medium and the dielectric constant of the whole system is the permittivity of water, $\varepsilon_C=\varepsilon(r)=\varepsilon_W=80$; model~2 (\model{step}) is the piecewise-constant dielectric model with $\varepsilon_C=2$ within the colloid and $\varepsilon(r)=\varepsilon_W=80$ outside; and model~3 (\model{ramp}) has an intermediate layer between the colloid and the bulk solvent, where the layer has thickness $2D_\text{ion} = \SI{0.9}{nm}$ and a linear transition dielectric permittivity, $\varepsilon(r)=(\varepsilon_W-\varepsilon_C)(r-R)/2D_\text{ion}+\varepsilon_C$ for $r\in[R, R+2D_\text{ion}]$.
For model~3 (\model{ramp}) in the HIM algorithm, the linear function $\varepsilon(r)$ in the transition layer is uniformly divided into 8 intervals ($L=10$) so that the approximate Green's function solution can be obtained. The results are verified to be accurate by using division refinement. For the MEMD algorithm, the linear function $\varepsilon(r)$ in the \model{ramp} model is interpolated onto a rectangular lattice of mesh size $a=\SI{0.104}{nm}$.

The simulation results are measured by the macroion-microion radial distribution function (RDF) and the integrated charge distribution function (ICDF). The RDF of each ion species is defined by,
\begin{equation}
g_\pm(r)=\frac{\langle N_\pm(r,r+\Delta r)\rangle}{\frac{4}{3}\pi[(r+\Delta r)^3-r^3]},
\end{equation}
which are normalized by $\int 4\pi r^2 [g_+(r)+g_-(r)]dr$, where $\langle N_\pm(r,r+\Delta r)\rangle$ is the mean particle number of the spherical shell between $r$ and $r+\Delta r$. The ICDF is the total charge within the sphere of radius less than $r$,
\begin{equation}
Q(r)=Q_M+Z_+e\langle N_+(a,r)\rangle+Z_-e\langle N_-(a,r)\rangle.
\end{equation}
The ICDF is useful for the computation of renormalized charges.

\section{Results and discussion}

\subsection{Structure of the EDL}

In the first group of simulated simulations, we set the bare colloidal charge to be $Z_C=\num{60}$, which corresponds to a surface charge density of $\SI{0.30}{e\per nm^2}$. We simulate a group of three systems. A number of monovalent salt ions, according to the desired salt concentration, are placed in these systems. Extra counterions are added to neutralize the macroion, such that the salt concentrations are $\num{20}$, $\num{40}$ and $\SI{60}{mM}$, respectively. To ensure that the salt concentration remains the same after ion condensation on the colloid, we measure the bulk salt concentration close to the cell boundary. We arrive at $\num{19.8}$, $\num{39.6}$, and $\SI{59.7}{mM}$ for the \model{ramp} model, and comparable results for the other two models. Additionally, this confirms that the size of the simulation cell is large enough to ensure a bulk concentration at the boundary. Therefore cell boundary effects or possible influences from the two different periodicities should not play a role in our results.

\begin{figure}[htbp!]
\subfloat[RDFs-mono][$c=\SI{20}{mM}$]{
    \centering
    \includegraphics[width=.9\linewidth]{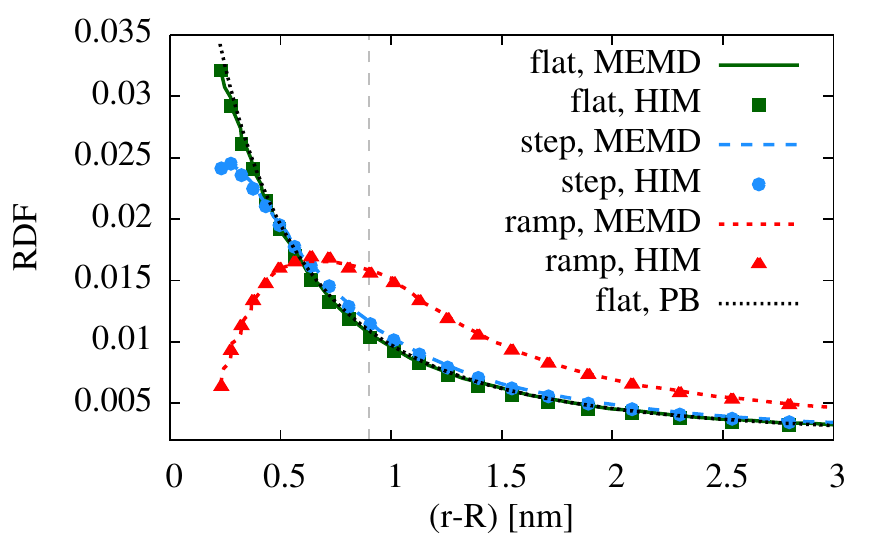}
    \label{fig:rdf-mono-20mM}
} \\
\subfloat[RDFs-mono][$c=\SI{40}{mM}$]{
    \centering
    \includegraphics[width=.9\linewidth]{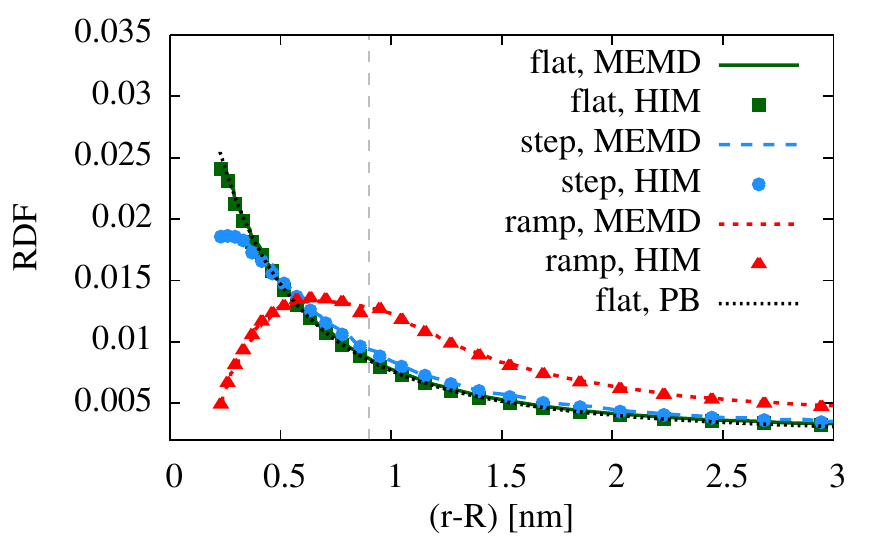}
    \label{fig:rdf-mono-40mM}
} \\
\subfloat[RDFs-mono][$c=\SI{60}{mM}$]{
    \centering
    \includegraphics[width=.9\linewidth]{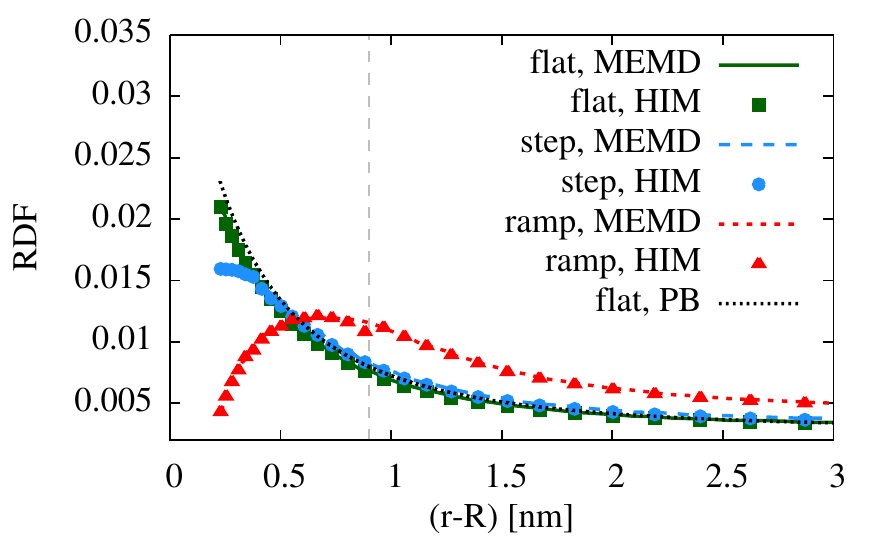}
    \label{fig:rdf-mono-60mM}
}
\caption{Counterion RDF curves of 1:1 salts calculated with three different dielectric models. The salt concentrations are $\SI{20}{mM}$, $\SI{40}{mM}$, and $\SI{60}{mM}$, in figure~\ref{fig:rdf-mono-20mM},~\ref{fig:rdf-mono-40mM}, and~\ref{fig:rdf-mono-60mM}, respectively. The bare surface charge is $Z_C=60$. For comparison, we have included the solution of a numerical Poisson-Boltzmann solver for the \model{flat}/\model{step} models, and the results overlap with our simulation. In the \model{step} model with a sharp dielectric contrast, the counterions are repelled from the colloid surface, something that is not included in the Poisson-Boltzmann solution, widening the double layer slightly. In the \model{ramp} model, the difference is a lot more pronounced, since ions can enter the region of lower permittivity, increasing the solvation energy and the electrostatic ion-ion repulsion. The crossover between the linear increase and the bulk is marked with a vertical line.} \label{fig:rdfs-mono}
\end{figure}

From the spherical concentration of ions, the radial distribution functions (RDFs) are calculated. The point of closest approach to the colloid is $r=\SI{4.225}{nm}$, the sum of colloid and ion radius. The counterion RDF curves are illustrated in Fig.~\ref{fig:rdfs-mono}. A difference between the two different algorithms is barely noticeable. This shows that our results can be trusted, in particular for the \model{ramp} model, for which the results to the best of our knowledge can not easily be confirmed by any other technique.

We observe, in agreement with Messina {\it et al.},\cite{messina01a} that the \model{flat} and \model{step} models show noticeable differences at close distances as well as in the far field, but both still result in monotonic counterion radial distributions. The \model{flat} model matches the theoretical prediction from a Poisson-Boltzmann solver very well. The \model{step} model slightly modifies the ion densities of the \model{flat} model near the colloid's surface, since the dielectric jump of the colloid-solvent interface creates an image charge repulsion on each mobile ion, reducing the counterion density near the surface.

The \model{ramp} model introduces a region of reduced dielectric permittivity around the colloid, into which the ions can enter. This significantly changes the structure of the EDL well beyond the region of varying permittivity between $\num{0}$ and $\SI{0.9}{nm}$ from the surface. The RDFs show a steep increase close to the surface and reach a maximum around $\SI{0.6}{nm}$. This can be explained by a solvation energy effect. The counterions are repelled from regions of low polarizability, which introduces a preferred motion towards the direction of $\nabla\varepsilon(r)$, as can be seen indirectly in equation~\eqref{eq:poisson}. Energetically, this can be linked to an increase in solvation energy when the ion enters a region of low dielectric permittivity, since the last term in equation~\eqref{eq:solvation-energy} dominates for small $\varepsilon$.
% Second, in the region of reduced dielectric permittivity, the same charge repulsion between the counterions is stronger than in a medium of high dielectric constant $\varepsilon=80$, and this pronounced ion-ion correlation introduces an electrostatic energy penalty for high salt concentrations close to the colloid surface.

\begin{figure}[htbp!]
    \centering
    \includegraphics[width=.9\linewidth]{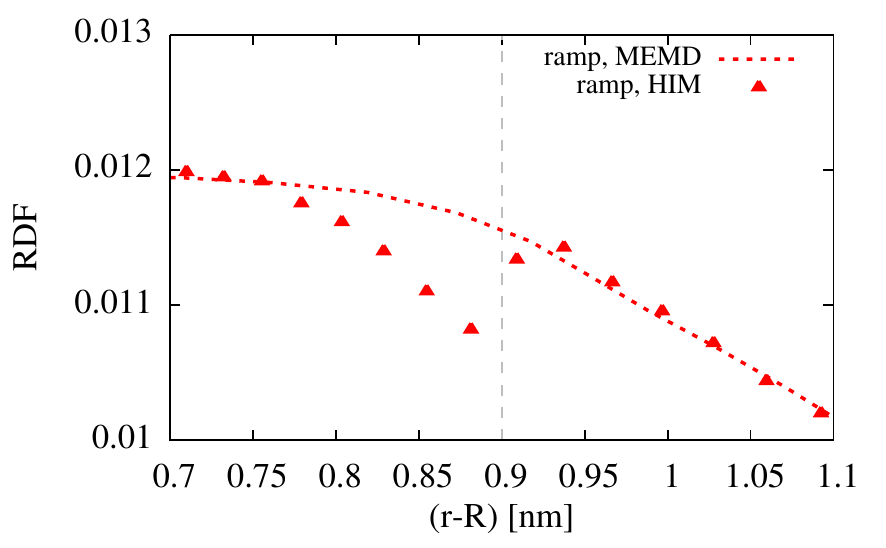}
    \caption{Closeup of the transition region between $\varepsilon(r)$ and $\varepsilon_W$ at $\SI{60}{mM}$ salt concentration. There is a defined kink with the HIM algorithm at the boundary due to the sharp jump of the permittivity gradient $\nabla\varepsilon$ at this point. This small effect is not present in the MEMD algorithm, since the interpolation of $\varepsilon$ onto a lattice corresponds to a smoothening of the dielectric function.}
    \label{fig:rdf-mc-kink}
\end{figure}

The RDFs of our MC simulations include a noticeable kink at the interface of linear interpolation $\varepsilon(r)$ and bulk, $\varepsilon_W$, meet at $\SI{0.9}{nm}$ (see more closely in Fig.~\ref{fig:rdf-mc-kink}). This is expected, since the gradient of the dielectric function $\varepsilon(r)$ has a sharp jump at this point. Ions in the region of increasing dielectric permittivity $(\nabla\varepsilon(r)>0)$ left of the jump will therefore be repelled from this jump $\nabla\varepsilon(\SI{0.9}{nm})=-\infty$. Since the dielectric function is also discretized into 8 intervals, the repulsion is not infinite but only occurs in the last interval from $\num{0.78}$ to $\SI{0.9}{nm}$. While this further reinforces the idea that sharp dielectric jumps in systems of freely moving charges are unphysical and will produce numerical artifacts, the influence in our case is very minor, as can be seen in Fig.~\ref{fig:rdf-mc-kink} by comparing to the MEMD simulation results.

In comparison, the EDL structures in Figs.~\ref{fig:rdf-mono-20mM},~\ref{fig:rdf-mono-40mM}, and~\ref{fig:rdf-mono-60mM} are not sensitive to the change of salt concentration. A threefold increase in the concentration (from $\num{20}$ to $\SI{60}{mM}$) only shows a small alteration of the local ionic distributions.

\subsection{Renormalized charge}

\begin{figure}[htbp!]
\subfloat[RDFs-Q][$Z_C=30$]{
    \centering
    \includegraphics[width=.9\linewidth]{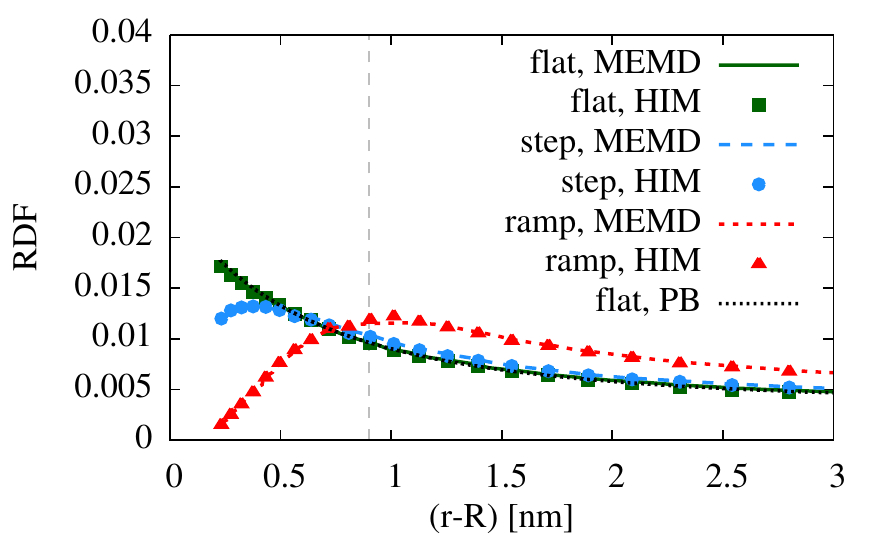}
    \label{fig:rdf-Q30}
} \\
\subfloat[RDFs-Q][$Z_C=60$]{
    \centering
    \includegraphics[width=.9\linewidth]{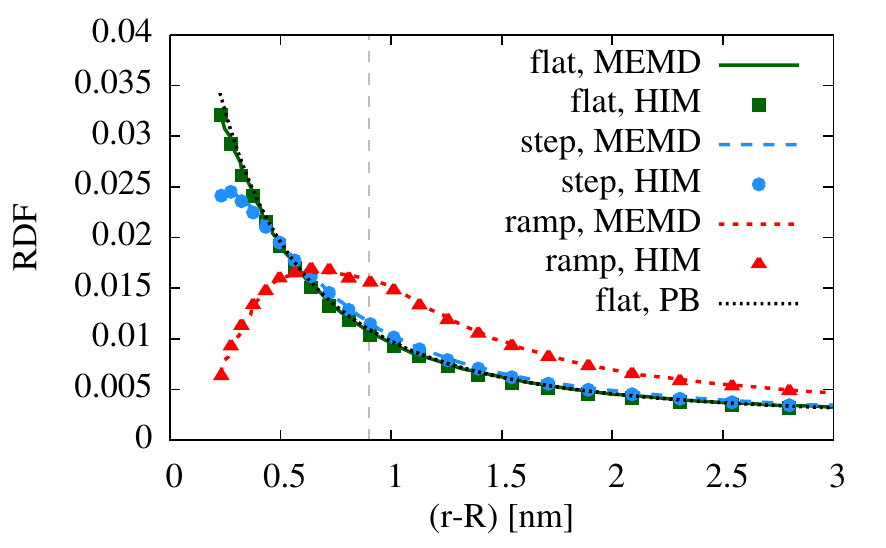}
    \label{fig:rdf-Q60}
} \\
\subfloat[RDFs-Q][$Z_C=90$]{
    \centering
    \includegraphics[width=.9\linewidth]{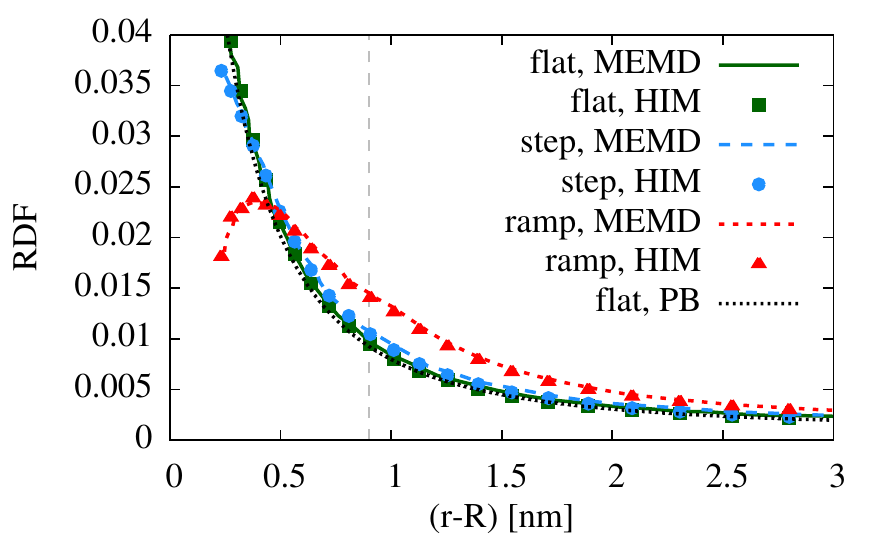}
    \label{fig:rdf-Q90}
}
\caption{Counterion RDF curves of 1:1 salts calculated with three different dielectric models. The salt concentration is $\SI{20}{mM}$, and the bare surface charges are $30$, $60$, and $90$, in Fig.~\ref{fig:rdf-Q30},~\ref{fig:rdf-Q60}, and~\ref{fig:rdf-Q90}, respectively. The crossover for the \model{ramp} model between the linear increase and the bulk is marked with a vertical line. For very low surface charges, the counterion distribution and therefore the Debye Layer are quite thick, reaching the counterion maximum only at $\SI{1}{nm}$ from the surface at $Z_C=30$. This leads to a comparably steep increase in the electrostatic potential in the far field.} \label{fig:rdfs-Q}
\end{figure}

\begin{figure}[htbp!]
    \centering
    \includegraphics[width=.9\linewidth]{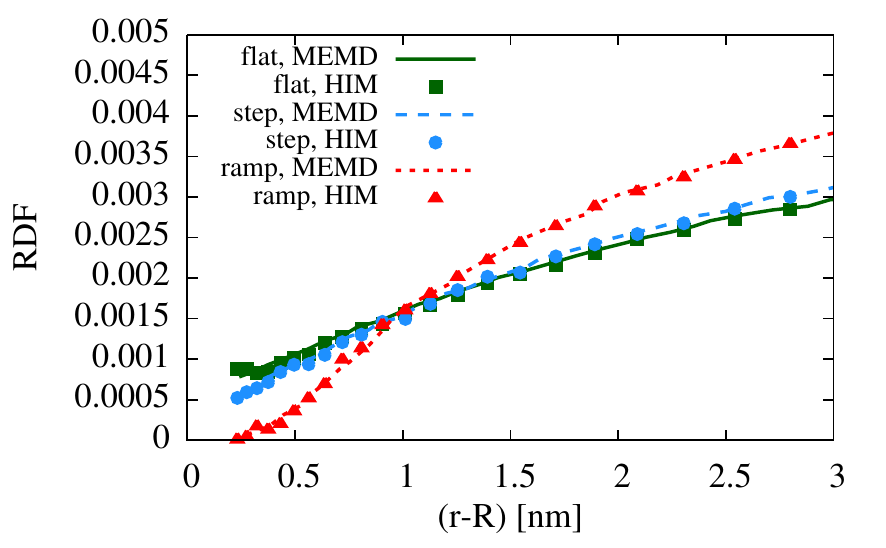}
\caption{Coion RDF curves of 1:1 salt calculated with three different dielectric models. The salt concentration is $\SI{20}{mM}$, and the bare surface charge is $\SI{30}{e}$. Where coions are able to enter the close vicinity of the colloid surface in the \model{flat} and \model{step} models, especially for low bare surface charges, the additional dielectric repulsion in the \model{ramp} model prevents this, forcing the coion density to zero at the surface.} \label{fig:rdfs-Q-coions}
\end{figure}

In the second group of simulated systems, we investigate the effect of inhomogeneous dielectric permittivity on the colloid's effective charge by varying the bare charge. We simulated the system described above at a constant salt concentration of $\SI{20}{mM}$, with increasing bare surface charge. The radial distribution functions for counterions and coions can be seen in Figs.~\ref{fig:rdfs-Q} and~\ref{fig:rdfs-Q-coions}.

%\begin{figure}[htbp!]
%\subfloat[RDFs-Q-coions][$Q=\SI{30}{e}$]{
%    \centering
%    \includegraphics[width=.9\linewidth]{rdf_monovalent_20mM_Q30_coions.pdf}
%    \label{fig:rdf-Q30-coions}
%} \\
%\subfloat[RDFs-Q-coions][$Q=\SI{60}{e}$]{
%    \centering
%    \includegraphics[width=.9\linewidth]{rdf_monovalent_20mM_Q60_coions.pdf}
%    \label{fig:rdf-Q60-coions}
%} \\
%\subfloat[RDFs-Q-coions][$Q=\SI{90}{e}$]{
%    \centering
%    \includegraphics[width=.9\linewidth]{rdf_monovalent_20mM_Q90_coions.pdf}
%    \label{fig:rdf-Q90-coions}
%}
%\caption{Coion RDF curves of 1:1 salts calculated with three different dielectric models. The salt concentration is $\SI{20}{mM}$, and the bare surface charges are $\SI{30}{e}$, $\SI{60}{e}$, and $\SI{90}{e}$, in Fig.~\ref{fig:rdf-Q30-coions},~\ref{fig:rdf-Q60-coions}, and~\ref{fig:rdf-Q90-coions}, respectively. Where coions are able to enter the close vicinity of the colloid surface in the \model{flat} and \model{step} models, especially for low bare surface charges, the additional dielectric repulsion in the \model{ramp} model prevents this, forcing the coion density to zero at the surface.} \label{fig:rdfs-Q-coions}
%\end{figure}

It is apparent from Fig.~\ref{fig:rdf-Q30} that the EDL is significantly thicker at very low surface charges, which leads to a steeper increase of the electrostatic potential in the far field. The structure of the EDL itself is also modified significantly, with counter- and coions being pushed away from the colloid surface very strongly, resulting in an ion density of almost zero for lower colloid charges. To further investigate this behavior, the so-called renormalized charge was calculated by fitting an extended Debye-H\"uckel formula~\eqref{eq:renormalizedcharge} below to the far field of the simulated ion distribution.

For charge renormalization, we use the concept from Alexander {\it et al.}~\cite{alexander84a} with an additional prefactor to include the colloid radius more prominently. Originally, the charge renormalization is applied to solve the nonlinear PB equation, which states that the electrostatic potential far from the surface can be described by the solution of the linearized PB equation, but with an effective renormalized charge. This renormalized charge replaces the bare charge of the colloid, which is reduced by screening effects due to the counterions. Then, the effective interaction between colloids can be calculated with the Debye-H\"uckel theory.~\cite{debye23a}

We calculate the renormalized charge through the following formula,
\begin{equation}
Z(r)=Z_\text{eff}\cdot\frac{1+\kappa r}{1+\kappa R}e^{-\kappa (r-R)},\label{eq:renormalizedcharge}
\end{equation}
which is adapted from the linearized PB equation as presented in the original publication.\cite{alexander84a}

%The value of this renormalized charge can hence be determined by fitting the DLVO formula
%%
%\begin{equation}
%V_\text{DLVO}(r) = Q_\text{eff}^2 \ell_B \left(\frac{e^{\kappa R}}{1+\kappa R}\right)^2 \frac{e^{\kappa r}}{r}
%\label{eq:dlvo-potential}
%\end{equation}
%%
%to the radial charge distribution. Here, $\kappa=\sqrt{4\pi \ell_B\sum_{k=1}^K n_k^0 Z_k^2},\label{debye}$ is the inverse Debye length for electrolytes with
%$K$ species of micro-ions, and $n_k^0$ and $Z_k$ are the concentration and valency of the $k$th species.

For dilute monovalent salts, the linearized PB equation holds in the region where the distance to the surface is larger than the Gouy-Chapman length,\cite{boroudjerdi05a}  $\ell_\mathrm{GC}=1/(2\pi Z \ell_B \sigma_S)$ where $\sigma_S=|Q_M|/4\pi R^2$ is the surface charge density. To ensure the validity of the fit, we discarded data points closer to the surface than $\num{2}\ell_\text{GC}$ and had very stable results from a least squares fit.

The simulation systems have distinct bare charges starting from $Z=-10$ to $-90$, which corresponds to a surface charge density varying from $\num{0.05}$ to $\SI{0.45}{e\per nm^2}$. The salt concentration is fixed at $\SI{20}{mM}$. The renormalized charge is calculated by fitting eq.~\eqref{eq:renormalizedcharge} via $\kappa$ and $Z_\text{eff}$ for $r-R$ from $\SI{1.5}{nm}$. In addition to the residuum of the least squares fit routine, the resulting values of $\kappa$ are a good control of the fit quality when compared to the calculated theoretical value, and they match very well.

\begin{figure}[htpb!]
\includegraphics[width=.9\linewidth]{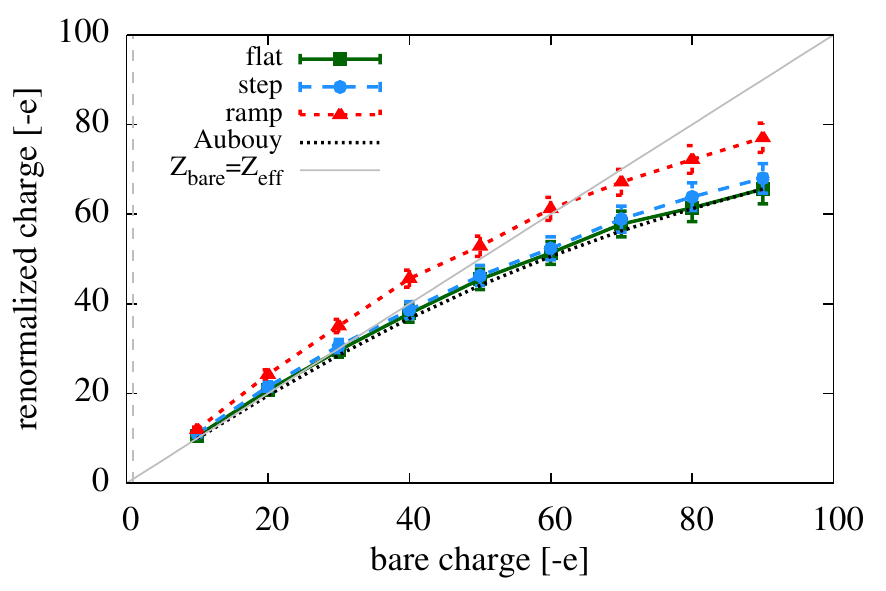}
%\vspace{1cm} \xu{Insert: The figure RCbar.eps}
%\vspace{1cm}
\caption{Renormalized charge versus the bare colloid charge for the three different dielectric models, as well as the results of a theoretical prediction, for a 1:1 electrolyte of $\SI{20}{mM}$. The \model{flat} and \model{step} models are very similar, and match the theory in the applicable region. The \model{ramp} model deviates from these significantly and demonstrates the fundamental influence of dielectric changes within the double layer. The counterions are repelled so strongly that, for low surface charges, the renormalized charge of the colloid is actually increased due the colloid appearing to have a larger effective radius.}\label{fig:renormalized-charges}
\end{figure}

The calculated effective charge with fit errors is plotted against the bare colloid charge in Fig.~\ref{fig:renormalized-charges}. For comparison, the theoretical prediction for a constant background permittivity as proposed by Aubouy {\it et al.}~\cite{aubouy03a} is included in the plot. It follows the formula
\begin{equation}
Z_\text{eff} = \frac{R}{\ell_B}\left[4\kappa R t_Q + 2\left(5-\frac{t_Q^4+3}{t_Q^2+1}\right)t_Q\right],
\end{equation}
where the bare charge is included via $t_Q = (\sqrt{1+x^2}-1)/x$
with the substitution $x :=  Z_\text{bare}\ell_B/2R(\kappa R + 1).$

In the renormalized charge, and therefore the far-field interaction, there is a minor difference between the primitive \model{flat} model and the \model{step} model with a sharp dielectric jump. We confirm the findings in ref.~\cite{messina02b} that the effective screening of the bare colloidal charge is slightly less effective, since the counterions are repelled from the colloid surface and the EDL is widened. The difference to the newly proposed \model{ramp} model however is far greater. It can already be seen in Fig.~\ref{fig:rdfs-mono} that the width of its EDL is significantly larger than those of the other two models. While the effective charge for the \model{flat} and \model{step} models ranges between $73\%$ and $99\%$ of the bare charge, the \model{ramp} model predicts much higher values between $86\%$ and $117\%$. This means that for surface charge densities of less than $\SI{0.3}{e\per nm^2}$, the effective charge of the colloid as defined in equation~\eqref{eq:renormalizedcharge} is higher than the bare charge. The EDL is widened significantly, and the counterion maximum is as far as $\SI{1}{nm}$ from the colloid surface, as can be seen in Fig.~\ref{fig:rdf-Q30}. This means that compared to Debye-H\"uckel theory, the electrostatic potential actually increases more steeply at far distances, which leads to an increase of the effective charge. This can be explained via an effective colloid radius. It is apparent from equation~\eqref{eq:renormalizedcharge} that the fitted effective charge will be larger if the colloid exhibits an effective radius that is larger than $R=\SI{4.225}{nm}$, as inserted into the formula. And with the counterions being pushed out very far, the far field will resemble that of a colloid with an increased effective radius. While effective charges higher than the bare colloid charge are unintuitive at first, they can occur with our definition of the charge renormalization and are meaningful. The \model{flat} model matches the theoretical curve closely as expected, since the theory does not include dielectric effects. All three models converge towards the line $Z_\text{bare} = Z_\text{eff}$ for very low surface charges, as expected.

From the data in Fig.~\ref{fig:renormalized-charges}, a shift $\Delta Z_\text{eff}$ in comparison to the \model{flat} model can be calculated and with equation~\eqref{eq:renormalizedcharge} interpreted as a change in effective colloid radius. With these calculations, the effective colloid charge corresponds to the radius being widened by $\SI{0.37}{nm}$, $\SI{0.29}{nm}$, and $\SI{0.19}{nm}$, for $Z_\text{bare}=30, 60,$ and $90$, respectively. This is a reasonable estimate when compared to the EDL structure in Fig.~\ref{fig:rdfs-Q}. This virtual effective colloid size is not an observable change in diameter, but a makeshift parameter that is usually included in the effective charge. In this case however, it explains why the calculated effective charge can be higher than the bare charge of the colloid.

\section{Conclusions}

The influence of charges in the dielectric background to the effective interaction between macroions has been subject to research for quite some time. Until now, it was not possible to directly include smooth changes of the dielectric permittivity in regions where charges can move freely. With the two recently introduced algorithms presented in this work, the long-range effect of these changes has been investigated. Both approaches give identical results.

We compared the electric double layer (EDL) structure for a charged colloid in electrolyte solution with three different models of the surrounding dielectric properties. The last of these models, a linear increase of the permittivity from the colloid surface, has not been accessible to coarse grained simulations up until now for the lack of a suitable electrostatic solver. The impact of this model is significant on the counterion profile within the EDL. It is even more pronounced in the difference between the renormalized charge calculations according to the Alexander prescription. The results indicate that a spatially varying permittivity $\varepsilon(r)$ likely plays a significant role in other biophysical systems, and that the presented algorithms should be applied more widely.

\section*{Acknowledgements}

FF and CH thank the DFG for support through the SFB 716 and the German Ministry of Science and Education (BMBF) for support under grant 01IH08001. ZX was supported by the Alexander von Humboldt foundation and the Natural Science Foundation of China (Grants No.: 11101276 and 91130012). Thank you to Prof. Xiangjun Xing and Stefan Kesselheim for helpful discussions.

%%%% Bibliography  %%%%%%%%%%
%%{unsrt}%{SIAM}% {abbrv} %{nature} %{plain} %{abbrv} %
%\bibliographystyle{elsart-num} %{elsart-num-sort}
%\bibliographystyle{ieeetr}
%\bibliography{dielectric_colloid}

\end{document}